\documentclass[twocolumn, nofootinbib, aps, prl, floats, floatfix, amsmath, amssymb, longbibliography, superscriptaddress, preprintnumbers]{revtex4-1} %
\usepackage[final]{graphicx}
\usepackage{hyperref}
\usepackage{amsmath}
\usepackage{bbm}
\usepackage{bm}
\usepackage{amsfonts}
\usepackage{amssymb}
\usepackage{latexsym}
\usepackage{graphicx}
\usepackage[english]{babel}
\usepackage{multirow}
\usepackage{float}
\usepackage{url}
\usepackage{slashed}
\usepackage{xcolor} 
\usepackage[utf8]{inputenc}
\usepackage{stmaryrd} 
\usepackage{enumitem}
\usepackage{hyperref}
\usepackage{cleveref}
\usepackage{siunitx}
\usepackage{verbatim}
\usepackage{braket}
\usepackage{mathrsfs}
\usepackage{makecell}

\usepackage{amsthm}

\theoremstyle{remark}

\newcommand{\be}{\begin{equation}}
\newcommand{\ee}{\end{equation}}
\newcommand{\ba}{\begin{array}}
\newcommand{\ea}{\end{array}}
\newcommand{\bea}{\begin{eqnarray}}
\newcommand{\eea}{\end{eqnarray}}

\newcommand{\bx}{{\bf x}}
\newcommand{\bq}{{\bf q}}

\newcommand{\besub}{\begin{subequations}}
\newcommand{\eesub}{\end{subequations}}

\newcommand{\nnn}{\nonumber \\}

\newcommand{\tphipp}{\delta\tilde\phi''}
\newcommand{\tphip}{\delta\tilde\phi'}
\newcommand{\tphi}{\delta\tilde\phi}
\newcommand{\mpl}{M_{\rm pl}}
\newcommand{\tPhi}{\tilde\Phi}
\newcommand{\tPsi}{\tilde\Psi}
\newcommand{\de}{\text{d}}
\newcommand{\e}{\text{e}}

\newcommand{\beq}{\begin{equation} \begin{aligned}}
\newcommand{\eeq}{\end{aligned} \end{equation}}

\newcommand{\hinf}{H_{\rm inf}}
\newcommand{\qp}{q_{\rm phys}}

\definecolor{darkerblue}{rgb}{0.2,0.2,0.5}
\definecolor{seagreen}{rgb}{0.180392,0.545098,0.341176}
\definecolor{smagenta}{rgb}{0.5,0.145098,0.341176}
\definecolor{deepblue}{rgb}{0,0,1}

\begin{document}

	\title{Phase transition during inflation and the gravitational wave signal at~pulsar~timing~arrays}
	
	\author{Haipeng An}
	\email{anhp@mail.tsinghua.edu.cn}
	\affiliation{Department of Physics, Tsinghua University, Beijing 100084, China}
	\affiliation{Center for High Energy Physics, Tsinghua University, Beijing 100084, China}
	\affiliation{Center for High Energy Physics, Peking University, Beijing 100871, China}	
	\affiliation{Frontier Science Center for Quantum Information, Beijing 100084, China}
	
	\author{Boye Su}
	\email{sby20@mails.tsinghua.edu.cn}
	\affiliation{Department of Physics, Tsinghua University, Beijing 100084, China}
	
	\author{Hanwen Tai}
	\email{hanwentai@uchicago.edu}
        \affiliation{Department of Physics, The University of Chicago, Chicago, IL 60637, USA}
	\affiliation{Enrico Fermi Institute, University of Chicago, Chicago, IL 60637, USA}

	\author{Lian-Tao Wang}
	\email{liantaow@uchicago.edu}
        \affiliation{Department of Physics, The University of Chicago, Chicago, IL 60637, USA}
	\affiliation{Enrico Fermi Institute, University of Chicago, Chicago, IL 60637, USA}
	\affiliation{Kavli Institute for Cosmological Physics, University of Chicago, Chicago, IL 60637, USA}
	
	\author{Chen Yang}
	\email{yangc18@mails.tsinghua.edu.cn}
	\affiliation{Department of Physics, Tsinghua University, Beijing 100084, China}

\begin{abstract}
	
The gravitational wave (GW) signal offers a promising window into the dynamics of the early universe. The recent results from the pulsar timing arrays (PTAs) could be the first glimpse of such new physics. In particular, they could point to new details during inflation, which can not be probed by other means. We explore the possibility that the new results could come from the secondary GWs sourced by curvature perturbations, generated by a first-order phase transition during inflation. Based on the results of a field-theoretic lattice simulation of the phase transition process, we show that the GW signal generated through this mechanism can account for the new results from the PTAs. We analyze the spectral shape of the signal in detail. Future observations can use such information to distinguish the GW signal considered here from other possible sources.

\end{abstract}
\maketitle

{\noindent}{\it \textbf{Introduction}---} Despite the enormous progress in our knowledge of the early universe, large gaps remain. The GW signal offers a new window into the early universe epochs and dynamics that cannot be probed by other means. 
PTA collaborations have recently released further evidence that the previously observed common-spectrum process exhibits the Hellings-Downs angular correlation~\cite{NANOGrav:2023hde,Antoniadis:2023utw,Zic:2023gta,Xu:2023wog,NANOGrav:2023gor,Antoniadis:2023rey,Reardon:2023gzh}. This finding indicates the existence of a gravitational wave background in the nano-Hz frequency range~\cite{Hellings:1983fr}. 

The GW signal in this frequency range can emerge from various phenomena, including low-scale phase transitions in the radiation domination (RD) era~\cite{Franciolini:2023wjm,Zu:2023olm,Han:2023olf,Fujikura:2023lkn,Athron:2023mer,Addazi:2023jvg,Jiang:2023qbm,Murai:2023gkv,Li:2023bxy,Xiao:2023dbb,Ghosh:2023aum,Abe:2023yrw,Du:2023qvj,Wu:2023hsa,Li:2023tdx,Cruz:2023lnq,DiBari:2023upq,Gouttenoire:2023bqy,Salvio:2023ynn}, supermassive black hole binaries (SMBHBs)~\cite{Shen:2023pan,Guo:2023hyp,Franciolini:2023pbf,Ellis:2023dgf,Broadhurst:2023tus,Huang:2023chx,Bi:2023tib,Zhang:2023lzt,Gouttenoire:2023nzr}, and topological defects like cosmic strings and domain walls~\cite{An:2023idh,Wang:2023len,Ellis:2023tsl,Kitajima:2023cek,Bai:2023cqj,Kitajima:2023vre,Lazarides:2023ksx,Depta:2023qst,Bian:2023dnv,Barman:2023fad,Li:2023tdx,Antusch:2023zjk,Babichev:2023pbf,Buchmuller:2023aus,Yamada:2023thl,Ge:2023rce,Zhang:2023nrs}. Simultaneously, this signal can also originate during inflation, approximately 15 e-folds after the cosmic microwave background (CMB) modes exit the horizon. Numerous mechanisms have been investigated in relation to this possibility~\cite{Vagnozzi:2023lwo,Oikonomou:2023qfz,Datta:2023vbs,Borah:2023sbc,Wang:2023ost,Gu:2023mmd,Chowdhury:2023opo,Niu:2023bsr,Ebadi:2023xhq,Yi:2023mbm,Figueroa:2023zhu,Unal:2023srk,Firouzjahi:2023lzg,Zhu:2023faa,Choudhury:2023kam,You:2023rmn,HosseiniMansoori:2023mqh,Cheung:2023ihl,Basilakos:2023xof,Jin:2023wri,Balaji:2023ehk,Bousder:2023ida,Li:2020cjj,Madge:2023cak}. Additionally, alternative new physics scenarios have been proposed as potential sources for the observed signal~\cite{Li:2023yaj,Yang:2023aak,Anchordoqui:2023tln,Konoplya:2023fmh,Geller:2023shn,Gelmini:2023kvo}. In this work, we focus on the possibility of the source being a first-order phase transition during inflation~\cite{Jiang:2015qor,An:2020fff,An:2022cce,Barir:2022kzo}. The process of bubble collision during the phase transition can generate GWs, referred to as primary GW signal, which is suppressed by $(\hinf/\beta)^5$ and $(L/\rho_{\rm inf})^2$, with $\hinf$ representing the Hubble expansion rate, $\beta$ denoting the phase transition rate, $L$ representing the latent heat density and $\rho_{\rm inf}$ being the total energy density of the universe during inflation. 
Meanwhile, the phase transition process serves as a source of curvature perturbations, that, after inflation, give rise to the so-called secondary GWs~\cite{Baumann:2007zm} (also see~\cite{Kohri:2018awv,Adshead:2021hnm}). 
Compared to the primary GWs, the secondary GWs can undergo natural enhancement via the slow-roll parameter, thus potentially accounting for the signals observed by the PTAs.

As pointed out in \cite{An:2020fff} and \cite{An:2022cce}, the plasma energy density during inflation is significantly smaller compared to latent heat. Therefore, the dominant source for GW and the induced curvature perturbation is from bubble collisions.  
In this work, we employ a field-theoretic simulation of the bubble nucleation and collision process to calculate the induced curvature perturbation. Based on the results, we predict both the strength and the spectral shape of the secondary GW signal. Without combining the data sets from different PTAs, we choose the results from the NANOGrav collaboration~\cite{NANOGrav:2023hde} as a benchmark for comparison. We show that both the size and the shape of the signal observed by NANOGrav~\cite{NANOGrav:2023hde,NANOGrav:2023gor} in the region with the frequency $f < {\rm (1 \ year)}^{-1}$ can be well fit by the secondary GWs generated through first-order phase transition occurring during inflation. 
 
\bigskip

\noindent{\it \textbf{The model}}--- In this work, we model a spectator sector with a single real scalar field, $\sigma$. 
The Lagrangian of the inflaton $\phi$ and the spectator $\sigma$  is 
\bea\label{eq:L}
{\cal L} =   - \frac{1}{2} g^{\mu\nu} \partial_\mu\phi \partial_\nu\phi - \frac{1}{2} g^{\mu\nu} \partial_\mu\sigma\partial_\nu\sigma - V(\phi,\sigma) \ .  
\eea
For the convenience of later discussions, we decompose $V(\phi,\sigma)$ as
\bea
V(\phi,\sigma) = V_0(\phi) + V_1(\phi,\sigma) \ ,
\eea
where $V_0(\phi) = V(\phi,0)$. 

We decompose $\phi$ into its homogeneous part $\phi_0$ and the perturbation $\delta\phi$ during inflation. The excursion of $\phi_0$ during inflation can be around the Planck scale. The crucial part in the Lagrangian (\ref{eq:L}) is that the mass of  $\sigma$ depends on $\phi_0$, given by
\bea
m^2_\sigma = c_m\phi_0^2-m^2 \ .
\eea
Consequently, the evolution of $\phi_0$ affects the shape of the potential and triggers a first-order phase transition. The general framework for this scenario is discussed in \cite{An:2020fff,An:2022cce}. Further details of the specific model utilized in our numerical simulation are presented in the supplemental material (SUPP). The timescale of the phase transition is determined by $\beta = - d S_4/d t$, where $S_4$ is the bounce action between the false and true vacuums, and $t$ is the physical time. 

For the class of models considered here, $\beta/\hinf \sim {\cal O}(10)$~\cite{An:2020fff,An:2022cce}.

Through this work, we adopt the Newtonian gauge. The metric components can be written as 
\bea
g_{00}\! = \!- a^2 (1 + 2 \Phi) , g_{0i}\! =\! 0 , g_{ij}\! =\! a^2 \!\left[\delta_{ij} (1 - 2 \Psi) + h_{ij} \right]. 
\label{eq:gauge}
\eea
During inflation, we have $a = - 1/\hinf\tau$ with $\tau$ the conformal time. During RD, we have $a = a_R^2 H_R \tau$, where $a_R$ and $H_R$ are the scale factor and the Hubble parameter at reheating. In this work, we assume de Sitter inflation with instantaneous reheating, which implies $H_R = \hinf$. 

\bigskip

\noindent{\it \textbf{Primary and secondary GWs}}--- In our setup, GWs can be copiously generated during and after inflation. In both cases, the GWs satisfy the differential equation 
\bea\label{eq:diff}
{h^{\rm TT}_{ij}}'' + 2 {\cal H} {h^{\rm TT}_{ij}}' - \nabla^2 h^{\rm TT}_{ij} = 16 \pi G {\cal T}_{ij} \ ,
\eea
where ``TT'' denotes the transverse and traceless components, ${\cal H} = a'/a$, and ${\cal T}_{ij}$ represents the source of GWs. 

During inflation, the main contribution to the GWs is from the TT components of the energy-momentum tensor composed of $\sigma$ and $\delta\phi$, where $\delta\phi$ is induced by the back reaction from the phase transition as discussed later. We call these contributions primary GWs. After being produced, the primary GWs will exit the horizon, and their field strength will be frozen to fixed values. The primary GWs will oscillate again once they reenter the horizon. 

In addition to the primary GWs, the phase transition also induces curvature perturbations leading to secondary GWs after inflation. In the Newtonian gauge described by Eq.~(\ref{eq:gauge}), the source ${\cal T}_{ij}$ consists of terms quadratic in $\Phi$~\cite{Baumann:2007zm,Kohri:2018awv}.

\bigskip

\noindent{\it \textbf{The induced curvature perturbations}}--- From the Lagrangian (\ref{eq:L}), we can derive the equation of motion for $\tphi$ (the Fourier transformation of $\delta\phi$),
\bea\label{eq:deltaphi}
\tphipp_\bq - \frac{2}{\tau} \tphip_\bq + \left(q^2 + \frac{1}{\hinf^2\tau^2}\frac{\partial^2 V_0}{\partial\phi_0^2}\right)\tphi_\bq = {\cal S}_\bq \ ,
\eea
where the source ${\cal S}_\bq$ is
\bea\label{eq:Sq}
 {\cal S}_\bq &=& - \frac{1}{\hinf^2 \tau^2} \left[ \frac{\partial V_1}{\partial\phi} \right]_\bq - \left\{\frac{2 \tPhi_\bq}{\hinf^2 \tau^2}  \left( \frac{\partial V_0}{\partial\phi_0} + \left[ \frac{\partial V_1}{\partial\phi} \right]_0 \right)\right. \nnn &&
 \left.+ \frac{\dot\phi_0}{\hinf\tau} \left( 3\tPsi'_\bq + \tPhi'_\bq\right) \right\} \ .
\eea
Here the symbol $[\dots]_\bq$ denotes the Fourier mode with comoving momentum $\bq$, and $\dot\phi_0$ denotes $d\phi_0/dt$. There are two source terms on the right-hand side of Eq.~(\ref{eq:deltaphi}), in which the first is from the direct interaction between $\phi$ and $\sigma$, whereas the second is purely gravitational. 
In the case of a polynomial interaction $c_m\phi^2 \sigma^2$, we have 
\bea\label{eq:V1p}
\left[\frac{\partial V_1}{\partial\phi}\right]_\bq =  c_m [\phi \sigma^2]_\bq \approx  c_m \phi_0 [\sigma^2]_\bq \ .
\eea
According to the Einstein equations, $\Phi$ and $\Psi$ satisfy the following differential equation
\bea\label{eq:psi}
\tilde\Psi'_\bq - \frac{\tilde\Phi_\bq}{\tau} = - 4 \pi G_N \left( \frac{ \dot\phi_0 \tphi_\bq}{\hinf\tau}  +  \left[ \frac{\partial_i}{\partial^2}(\sigma' \partial_i\sigma) \right]_\bq\right) \ .
\eea
From energy-momentum conservation, we have
\bea\label{eq:pi}
\tilde\Phi_\bq - \tilde\Psi_\bq = - 8\pi G_N \tilde\pi^S_\bq / (\hinf^2 \tau^2) \ ,
\eea
where $\pi^S$ represents the anisotropic inertia
\bea\label{eq:piq}
\tilde\pi^S_\bq = -\frac{3}{2}\hinf^2\tau^2 q_iq_jq^{-4}\left[(\partial_i\sigma \partial_j \sigma)^{\rm TL} \right]_\bq \ ,
\eea
where the superscript TL refers to the traceless part.

After the phase transition, the universe returns to single-field inflation. Thus, the gauge-invariant quantity,
\bea\label{eq:zeta}
\zeta_\bq = - \tPsi_\bq - \frac{\hinf \tphi_\bq}{\dot\phi_0} \ ,
\eea 
is conserved when evolving outside the horizon.


\bigskip

\begin{figure}[h!]
	\centering
	\includegraphics[width=1\linewidth]{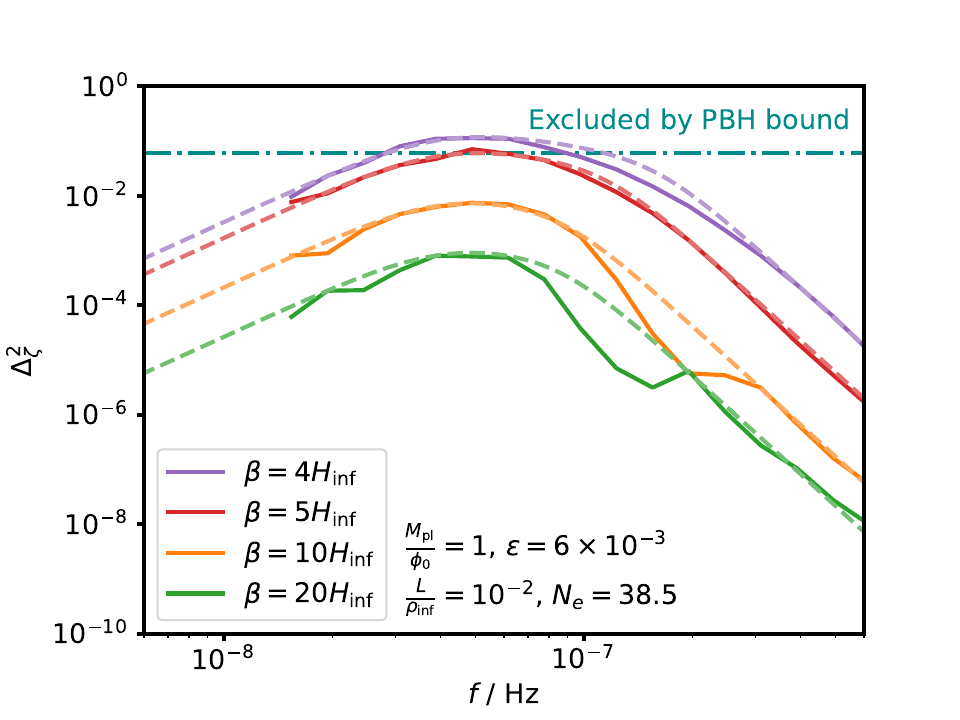}
	\caption{Power spectrum of the induced curvature perturbation, $\Delta^2_\zeta$ for different choices of parameters. The solid curves are the results from numerical simulation, and the dashed curves are based on the empirical formula Eq.~(\ref{eq:DeltaFormula}). The wiggles in the curves for $\beta/H_{\mathrm{inf}} = 10$ and 20 are the remnants of the oscillatory pattern in the integrand of (\ref{eq:zetaq}), which also leads to the oscillatory pattern in the primary GW spectrum as discussed in \cite{An:2020fff,An:2022cce}. Large curvature perturbations can result in primordial black hole (PBH) production, and the relevant constraint is indicated by the dash-dotted line~\cite{Byrnes:2018txb}. 
 }\label{fig:Pzeta}
\end{figure}

\noindent{\it \textbf{Power spectrum of $\zeta$}}--- In this work, we use a $1000\times1000\times1000$ lattice to simulate the phase transition process in de Sitter space and numerically solve Eqs.~(\ref{eq:deltaphi}), (\ref{eq:psi}) and (\ref{eq:pi}) to calculate the various contributions to  $\zeta_\bq$. As discussed in the appendix of \cite{An:2020fff}, the typical values for $\beta/H_{\rm inf}$ is ${\cal O}(10)$.
The solid curves in Fig.~\ref{fig:Pzeta} represent the numerical results of $\Delta^2_\zeta$, the spectrum of the induced curvature perturbation for $\beta/H_{\mathrm{inf}} = 4$, 5, 10 and 20. $\Delta^2_\zeta$ is defined as
\bea
\Delta^2_\zeta(q) = \frac{q^3}{2\pi^2} P_\zeta(q) = \frac{q^3}{2\pi^2} \langle\zeta_\bq \zeta^*_{\bq'}\rangle' \ , 
\eea
where $\langle\cdots\rangle'$ denotes the correlation function without the delta function. 
Fig.~\ref{fig:Pzeta} shows that $\Delta^2_\zeta$ grows as $q^3$ in the IR region (to the left of the peak) and drops as $q^{-6}$ in the UV region (far to the right). By comparing the peak values of the curves for different $\beta/H_{\mathrm{inf}}$ in Fig.~\ref{fig:Pzeta}, it can be concluded that $\Delta^2_\zeta \propto (\hinf/\beta)^3$. 

To provide further insights, an approximate formula for $\Delta^2_\zeta$ will be derived. 
Several e-folds after the phase transition, the $\Phi$ contribution to $\zeta$ becomes negligible as its direct source from $\sigma$ quickly redshifts away. 

Hence, using the Green's function method and Eq.~(\ref{eq:deltaphi}), we obtain

\bea\label{eq:zetaq}
\zeta_\bq = - \frac{\hinf}{\dot\phi_0 q^2} \int_{-\infty}^0  \frac{d\tau'}{\tau'} \left( \cos q\tau' - \frac{\sin q\tau'}{q\tau'} \right) {\cal S}_\bq(\tau') \ .
\eea
As discussed in SUPP, in the parameter space of interest, the gravitational-induced contribution to the curvature perturbation is smaller than the direct contribution. Thus, in the qualitative analysis, we focus on the contribution from the direct source (the first term in ${\cal S}_\bq$).

According to Eqs.~(\ref{eq:V1p}) and (\ref{eq:zetaq}), $\Delta^2_\zeta$ is inversely proportional to $\epsilon\equiv\dot\phi_0^2/(2 \hinf^2 \mpl^2)$, the slow-roll parameter after the phase transition. Assuming that the rolling of the scalar fields is predominantly in the direction of $\phi_0$ after the phase transition, $\zeta_\bq$ can be estimated as
\bea\label{eq:20}
\zeta_\bq \approx  \frac{\hinf}{ \dot\phi_0 q^2}\! \int\! \frac{d\tau'}{ \tau'} \!\left(\! \cos q\tau'\! -\! \frac{\sin q\tau'}{q\tau'}\! \right)\! \frac{c_m \phi_0 [\sigma^2(\tau')]_\bq}{\hinf^2 \tau'^2} \ . 
\eea
For a first-order phase transition to complete, we require its duration $\sim\beta^{-1} <  \hinf^{-1}$. However, even after the phase transition, the $\sigma$ field continues to oscillate and produce $\zeta$. Since $\beta < m_\sigma$, these oscillations resemble matter-like behavior and redshift as $a^{-3}$. Thus, the $\zeta$ production diminishes within a few e-folds after the phase transition. Fig.~S2 in SUPP shows that majority of induced curvature perturbations occur between one and two e-folds after the phase transition begins. 

For modes with $q_{\rm phys} < \hinf$, or equivalently $q \tau'  < 1$, a Taylor expansion of the cosine and sine functions in the integrand of Eq.~(\ref{eq:20}) can be performed, resulting in 
\bea\label{eq:21}
\zeta_\bq \approx  \frac{1 }{3 \dot\phi_0} \int dt'  c_m \phi_0 [\sigma^2(\tau')]_\bq \ ,
\eea
where $d t' = a(\tau') d\tau' $. 
Considering that the typical scale of the bubble size is $\beta^{-1}$ and $\hinf < \beta$, the correlation $\langle [\sigma^2]_\bq [\sigma^2]_{\bq'} \rangle'$ is expected to be insensitive to $\bq$. Since the term $c_m\phi_0^2 \sigma^2$ triggers the phase transition, we also expect $c_m\phi_0^2 \sigma^2 \sim L$. Hence, we have,  
\bea\label{eq:22}
\int \!d t'\! \int\! dt''\! c_m^2\! \phi_0^2 \langle [\sigma^2(\tau')]_\bq [\sigma^2(\tau'')]_{\bq'}\rangle' \sim \frac{L^2}{\hinf^2 a_\star^6}\! \left(\!\frac{2\pi}{\beta}\!\right)^3  \!\ ,
\eea
where $a_\star$ is the scale factor at the time of the phase transition.
The factor $\hinf^{-2}$ arises due to the integration over the physical time duration, and the factor $(2\pi/\beta)^{3}$ is from dimensional analysis. The factor $a_\star^6$ appears in the denominator because the Fourier transformation is performed in comoving space. Combining Eqs. (\ref{eq:21}) and (\ref{eq:22}), in the region $q_{\rm phys} < \hinf$, we have, 
\bea\label{eq:DeltaIR}
\Delta^{2{\rm(IR)}}_\zeta(q) = A_{\rm ref} \left( \frac{q_{\rm phys}}{\hinf} \right)^3 \ ,
\eea
where 
\bea
A_{\rm ref} = \frac{{\cal A}}{\epsilon}\left(\frac{\mpl}{\phi_0}\right)^2 \left(\frac{\hinf}{\beta}\right)^3 \left(\frac{L}{\rho_{\mathrm{inf}}}\right)^2 \ .
\eea
The numerical factor ${\cal A}$ is determined by the specific phase transition model. Numerical simulations show that ${\cal A}\approx 24$ for the model used in this work. 
Eq.~(\ref{eq:DeltaIR}) explains the IR behavior, the $(\hinf/\beta)^3$ dependence, and the $(L/\rho_{\mathrm{inf}})^2$ dependence of $\Delta^2_\zeta$ depicted in Fig.~\ref{fig:Pzeta}. 

The $q^3$ growth of $\Delta^2_\zeta$ stops when $q_{\rm phys}$ reaches $H$. In the region where $q_{\rm phys}\gtrsim\hinf$, the sine and cosine terms in (\ref{eq:20}) are of order one, while $\langle [\sigma^2]_\bq [\sigma^2]_{\bq'}^*\rangle'$ remains insensitive to changes of $q$. By counting the power of $q$, it can be concluded that $\Delta^2_\zeta$ drops as $q^{-1}$ in this region. 

As shown in SUPP, the production of the curvature perturbation lasts for a few e-folds, during which the typical physical momentum of the bubbles undergoes redshifting. As a result, the correlation $\langle [\sigma^2]_\bq [\sigma^2]_{\bq'}^*\rangle'$ drops significantly as $q_{\rm phys}$ become larger than a value between $H$ and $\beta$. This behavior leads to the drop of $\Delta_\zeta^2$ as $q^{-6}$ in the UV region, as shown in Fig.~\ref{fig:Pzeta}. 

In summary, we arrive at an empirical formula for $\Delta_\zeta^2$, 
\bea\label{eq:DeltaFormula}
{\Delta^2}^{(\rm emp)}_\zeta(q) = A_{\rm ref} {\cal F}\left( \frac{q_{\rm phys}}{\hinf} \right) \ ,
\eea
where 
\bea
{\cal F}(x) &=& \frac{x^3}{1 + (\alpha_1 x)^4 + (\alpha_2 x)^{9}} \  \ .
\eea
Numerical results give that $\alpha_1 = 0.31$, while $\alpha_2$ mildly depends on $\beta/\hinf$ and equals $0.143$, $0.17$, $0.2$, and $0.2$ for $\beta/\hinf = 4, 5, 10$ and 20.

In a general inflation model, $\phi_0$ and $\epsilon$ can be independent parameters. 
For single field inflation, if we consider the case that $\partial V_1/\partial\phi$ dominates the evolution of $\phi_0$ after the phase transition, we have 
\bea
\epsilon = \frac{\dot\phi_0^2}{2\mpl^2 \hinf^2} \sim \frac{(V_1')^2}{6\rho_{\mathrm{inf}} \hinf^2} \ ,
\eea
and, assuming $V_1$ is a polynomial in $\phi$, 
\bea
\frac{\mpl^2}{\phi_0^2} \sim \left( \frac{\mpl}{V_1/V_1'} \right)^2 \sim \left( \frac{\mpl}{L/V_1'} \right)^2 \ .
\eea
Thus, without fine-tuning, the peak value of $\Delta^2_\zeta$ is
\bea
\Delta_\zeta^2(q) \approx 3.6 \times \left(\frac{\hinf}{\beta}\right)^3 {\cal F}\left(\frac{\qp}{\hinf}\right)  \ ,
\eea
suggesting that for $\hinf/\beta\sim0.1$, a peak value of $\Delta_\zeta^2$ around 0.01 can be naturally expected. 

\begin{figure}
	\centering
	\includegraphics[width=1\linewidth]{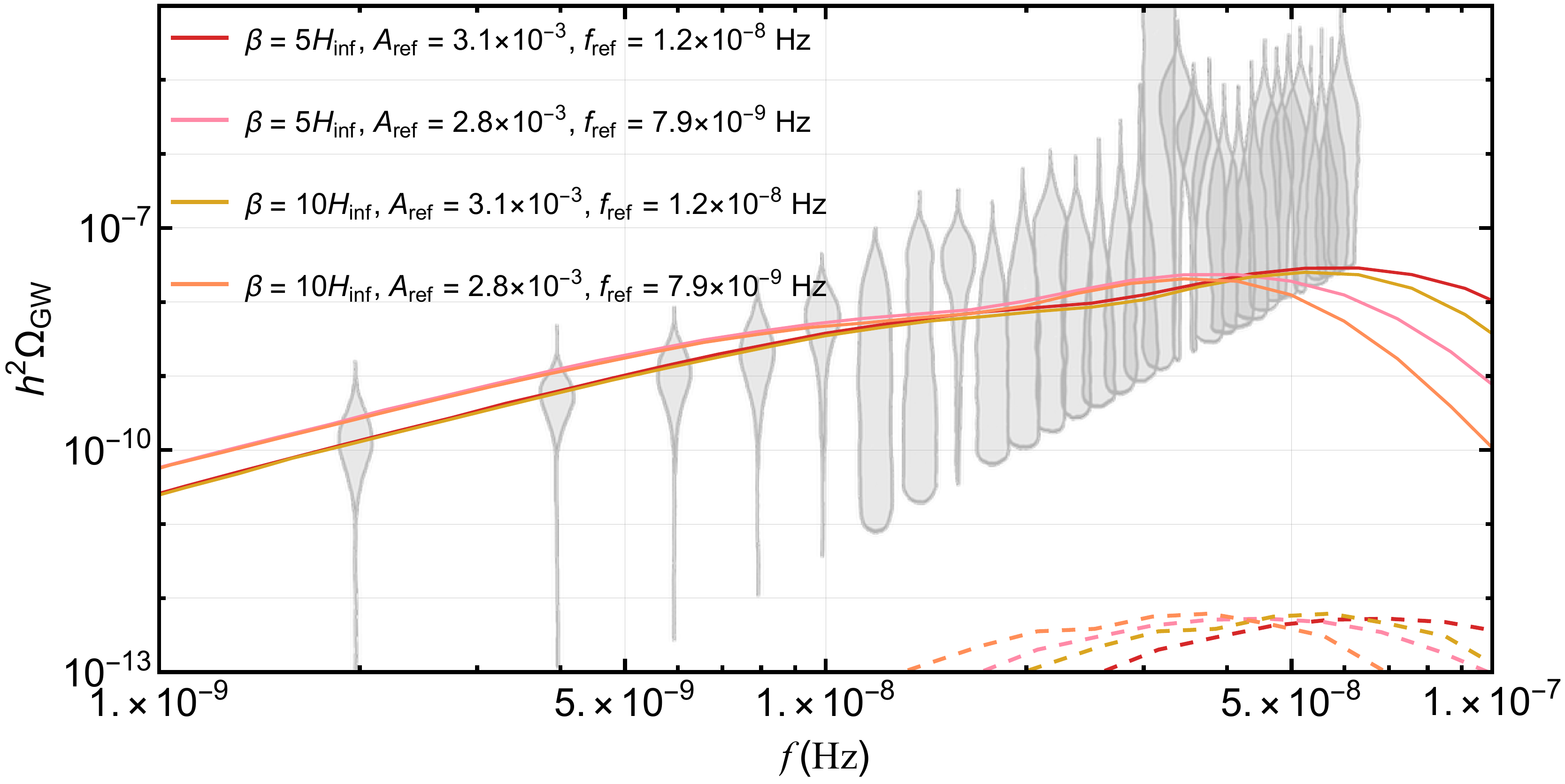}
	\caption{The differential spectra of secondary GWs induced by first-order phase transitions during inflation for different parameters as shown in the plot. The gray violins show the periodogram for an HD-correlated free spectral process from~\cite{NANOGrav:2023hvm}. Four sets of the model parameters are shown as examples,  which match the data collected by the NANOGrav collaboration in terms of the amplitude and spectral shape, particularly in the region $f< 3 \times 10^{-8}$ Hz. The corresponding primary GW spectra are represented by the dashed curves. In the parameter region of interest, the magnitude of the primary GWs is smaller than the secondary GWs by a few orders of magnitude.  
 }\label{fig:GW2}
\end{figure}

\bigskip

\begin{figure}
	\centering
	\includegraphics[width=0.95\linewidth]{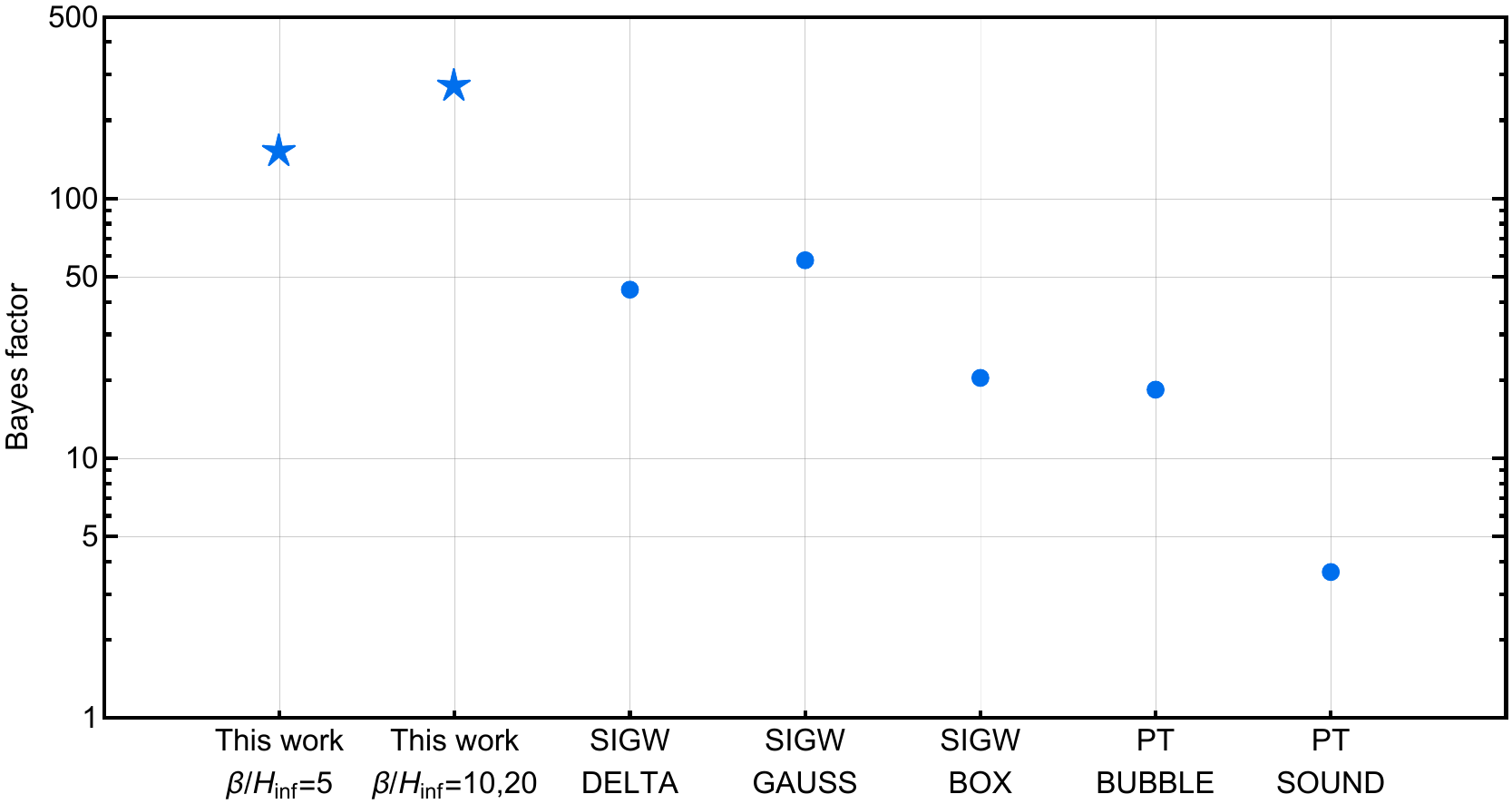}
	\caption{The Bayes factors for the model comparisons between the new-physics interpretations of the signal and the interpretation  based solely on SMBHBs. The stars denote the Bayes factors for the model considered in this work. Other results labeled as ``SIGW DELTA'', ``SIGW GAUSS'', ``SIGW BOX'', ``PT BUBBLE'', and ``PT SOUND'' are the benchmark new physics models studied in~\cite{NANOGrav:2023hvm}. SIGW and PT stand for scalar-induced GWs and phase transition, respectively.
 }\label{fig:bayes}
\end{figure}

\begin{figure}
	\centering
	\includegraphics[width=0.9\linewidth]{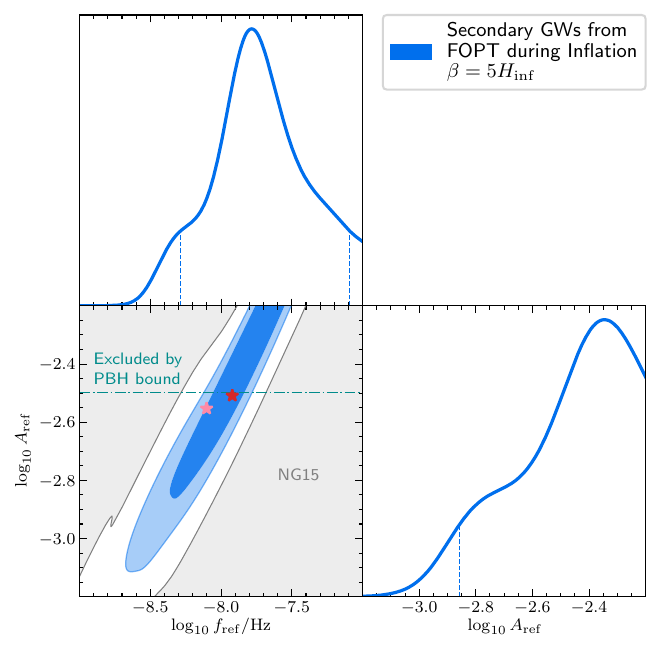}
	\caption{The reconstructed posterior distributions for the parameters, $A_{\rm ref}$ and $f_{\rm ref}$ for $\beta=5\hinf$. The upper left and lower right panels display the 1D marginalized distributions with the 68\% Bayesian credible intervals, while the lower left panel depicts the 68\% (darker) and the 95\% (lighter) Bayesian credible regions in the 2D posterior distribution. The two benchmark stars correspond to the two curves with same color shown in Fig.~\ref{fig:GW2}. The PBH bound on the curvature perturbation is also shown by the dash-dotted curve. }\label{fig:posteriors1}
\end{figure}

\begin{figure}
	\centering
	\includegraphics[width=0.9\linewidth]{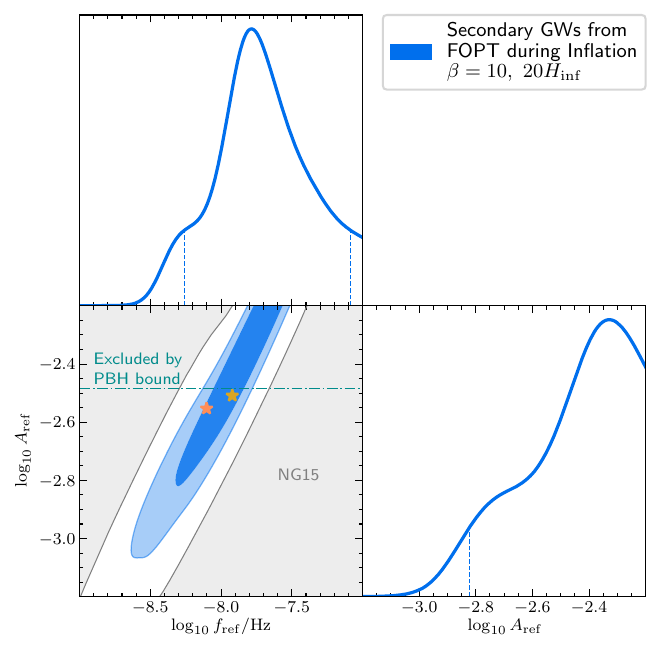}
	\caption{Same as in Fig.~\ref{fig:posteriors1} but for $\beta=10,\ 20\hinf$.}\label{fig:posteriors2}
\end{figure}

\noindent{\it \textbf{Production of secondary GWs}}--- To produce the GWs to account for the PTA results, the curvature perturbations need to reenter the horizon well before the matter-radiation equality. 

Following the procedure outlined in~\cite{Kohri:2018awv}, the spectrum function for the secondary GWs can be derived as

\bea
\Omega^{(2)}_{\rm GW}(f) = \Omega_R A_{\rm ref}^2 {\cal F}_2\left(\frac{q_{\rm phys}}{\hinf}\right),
\eea
where $\Omega_R$ is the radiation energy density of the universe. The form factor ${\cal F}_2$ collects transfer functions and Green's functions. As shown in SUPP, ${\cal F}_2(x)$ peaks around $x=5$, and its peak value is about 200. In the IR region, we have
\bea
{\cal F}^{\rm IR}_2(x) \approx x^3 \left( \frac{6}{5}\log^2 x + \frac{16}{25}\log x + \frac{28}{125} \right) \ .
\eea
The logarithmic structure in ${\cal F}_2$ slows down the rise of the spectrum, allowing for a better fit to the IR part of the NANOGrav observation data as depicted in Fig.~\ref{fig:GW2}. 

The relation between the comoving momentum $q$ and today's frequency $f$ is
\bea
f = \frac{q}{2\pi a_0} = f_{\rm ref} \times\frac{q_{\rm phys}}{\hinf} \ ,
\eea
where 
\bea
f_{\rm ref} = 10^{-9}~{\rm Hz} \times e^{40 - N_e}  \left(\frac{\hinf}{10^{14}~ {\rm GeV}}\right)^{1/2}\ ,
\eea
$N_e$ is the number of e-folds the phase transition happened before the end of inflation. Here we assume the reheating process finished within one e-fold. Thus, if the phase transition is responsible for the PTA signals, it happened about 40 e-folds before the end of inflation. 

Utilizing the Bayesian analysis developed by the NANOGrav collaboration~\cite{Mitridate:2023oar} (also described in SUPP), we present the Bayes factor for the model comparisons between our work and the interpretation in terms of SMBHBs in Fig.~\ref{fig:bayes} accompanied with the benchmark model chosen by the NANOGrav group~\cite{NANOGrav:2023hvm}, where we can see that the Bayes factor of our model is significantly higher than the benchmark models in~\cite{NANOGrav:2023hvm}. Fig.~\ref{fig:posteriors1} and Fig.~\ref{fig:posteriors2} display the reconstructed posterior distributions of our scenario, indicating that the benchmarks presented in Fig.~\ref{fig:GW2} fall within the 1$\sigma$ contour. It is interesting to note that the scenario for $\beta/\hinf=20$ is exactly the same as that for $\beta/\hinf=10$ in Fig.~\ref{fig:bayes} and Fig.~\ref{fig:posteriors2}, due to the fact that they share both form factors $\mathcal{F}\left(\frac{q_\text{phys}}{\hinf}\right)$ and $\mathcal{F}_2\left(\frac{q_\text{phys}}{\hinf}\right)$. Therefore, the spectra of secondary GWs for $\beta/\hinf=20$ completely overlap with those for $\beta/\hinf=10$, but with different primary GWs.

\bigskip

\noindent{\it {\textbf{Summary and outlook}}}--- The main results of this work is shown in Fig.~\ref{fig:GW2}. As a benchmark, we compare them with data from the NANOGrav collaboration. The observations from the other PTA collaborations are in broad agreement. The secondary GW signal considered in this work can provide a comparable magnitude to account for the observed data and fit the spectral shape, particularly in the region $f< 3 \times 10^{-8}$ Hz. 
Hence, it is concluded that the GW production mechanism investigated in this paper offers a promising explanation for the observations made by PTA collaborations. 

The observed signal in the higher frequency range $f > 3 \times 10^{-8}$ Hz seems to indicate an even higher amplitude. Further data and combining the data from all PTA collaborations can shed more light on this region. In the proposed scenario, it is possible, in principle, to consider slightly later phase transitions (smaller $N_e$) to shift the signal peak towards higher frequencies. Other parameters such as $L$, $\epsilon$, and $\beta$, can also be adjusted to achieve a higher amplitude. However, as already evident in Fig.~\ref{fig:Pzeta}, higher amplitudes of curvature perturbations will inevitably lead to copious production of primordial black holes and may be in tension with observations. This limitation applies to any mechanism of secondary GW production.

Significant PBH production is expected in the region with large curvature perturbations considered in this work, even if they have not been excluded yet. This could offer a correlated signal to verify the secondary GW production mechanism. A detailed investigation of this question is left for future work.

\bigskip

\bigskip

\noindent \textit{\textbf{Acknowledgment}}---
The work of HA is supported in part by the National Key R\&D Program of China under Grants No. 2021YFC2203100 and No. 2017YFA0402204, the NSFC under Grant No. 11975134, and the Tsinghua University Dushi Program No. 53120200422. The work of LTW is supported by the DOE grant DE-SC0013642.

\bibliography{ref}
\bibliographystyle{utphys}

\appendix

\section{Details of our numerical simulation}

Here, we present a comprehensive description of our numerical simulation, especially focusing on the evolution of the spectator field $\sigma$ and the inflaton field $\phi$ throughout the first-order phase transition. This section is divided into two subsections. Subsection~\ref{1A} provides an overview of the phase transition model employed in our numerical simulation. Subsection~\ref{1B} offers a detailed explanation of the lattice simulation method utilized in this study.

\subsection{The phase transition model}\label{1A}

In our simulation, the potential in the spectator sector is \begin{equation}\label{V}
	V_1\left(\phi,\sigma\right)=-\frac{1}{2}(m^2-c_m \phi^2)\sigma^2+\frac{1}{3}c_3\sigma^3+\frac{1}{4}\lambda\sigma^4\ .
\end{equation}
$V_1$ has two non-degenerate local minima, $\sigma_{\text{fl}}$ and $\sigma_{\text{tr}}$. In the earlier time of inflation, when $c_m \phi_0^2 > m^2+2c_3^2/\left(9\lambda\right)$, $\sigma = 0$ is the preferred vacuum. During the slow-roll inflation, $\phi_0$ becomes smaller, the phase transition happens when $c_m \phi_0^2 < m^2+2c_3^2/\left(9\lambda\right)$. 

The cubic term in the potential $V_1$ provides a barrier between the true and false vacua, and thus the phase transition is first-order. The bubble nucleation process is described by the bounce solution, $\sigma_{\text{b}}$ \cite{Coleman:1977py,Callan:1977pt}, which satisfies the Euclidean field equation,
\begin{equation}\label{eq:bounce}
	\frac{\de^2\sigma_{\text{b}}}{\de r^2}+\frac{3}{r}\frac{\de\sigma_{\text{b}}}{\de r}=\frac{\de V_1}{\de\sigma_{\text{b}}}\ ,
\end{equation}
where in Euclidean space $r = (t^2 + \bx^2)^{1/2}$ and $dr^2 = dt^2 + d\bx^2$.
In Eq.~(\ref{eq:bounce}), the Hubble expansion is ignored, since typical energy scale of the spectator sector in this study is much larger than $\hinf$.  
The boundary conditions for the Euclidean equation of motion (\ref{eq:bounce}) are
\begin{equation}
	\sigma_{\text{b}}\left(\infty\right)=\sigma_{\text{fl}},\ \left.\frac{d\sigma_{\text{b}}}{d r}\right|_{r=0}=0\ ,
\end{equation}
and then the bubble nucleation rate per unit physical volume can be written as  
\begin{equation}
	\frac{\Gamma}{V_{\rm phys}}\sim m^4\e^{-S_{\text{b}}},
\end{equation}
where the Euclean bounce action $S_{\rm b}$ is 
\bea\label{eq:Sb}
S_{\text{b}}&=&2\pi^2\int_0^{\infty}d rr^3 \nnn
&&\left[\frac{1}{2}\left(\frac{d\sigma_{\text{b}}}{dr}\right)^2+V_1\left(\phi_0,\sigma_{\text{b}}\right)-V_1\left(\phi_0,\sigma_{\text{fl}}\right)\right] \ .
\eea
The initial configuration of the bubbles in Minkowski space are then given by the analytical continuation of bounce solution to the Minkowski space, $\sigma_{\text{b}}\left(r\right)\rightarrow\sigma_{\text{b}}\left(\sqrt{a^2\left(\mathbf{x}^2-\tau^2\right)}\right)$. 
Then we use the classical field equation,
\begin{equation}\label{eq:sigma}
	\sigma''+2\mathcal{H}\sigma'-\nabla^2\sigma+a^2\frac{d V_1}{d\sigma}=0\ ,
\end{equation}
to calculate the evolution of the $\sigma$ field. 

\subsection{Lattice simulation}\label{1B}

We discretize the space using a cubic lattice with $N$ grids per spatial dimension. The $N^3$ points are labeled as
\begin{equation}
	\mathbf{n}=\left(n_1,n_2,n_3\right)\ ,\ \text{with}\ n_i=0,1,\dots,N-1,\ i=1,2,3\ .
\end{equation}
In the simulation, a field $f\left(\mathbf{x}\right)$ defined on the continuum space with comoving coordinate $\bx$, is converted to a field $f\left(\mathbf{n}\right)$ defined on the lattice, with the condition that $f\left(\mathbf{x}\right)$ is calculated at $\mathbf{x}=\mathbf{n}\delta x$. Note that in our simulation, the lattice is defined in coordinate system. For convenience, we set $m=1$ hereafter. In this unit system, the Hubble constant $H_{\rm inf}$ has a value of $0.01$. The physical size of the lattice spacing $\delta x$ is set to $0.6$ at the beginning of the simulation. We have verified that this value is sufficiently small to provide accurate secondary GW signals by comparing the results obtained with $\delta x = 0.6$ to those obtained with $\delta x = 0.3$, while maintaining the same total volume. 

In our simulation, we consider $N=1183$, resulting in a total space size of $7H_{\rm inf}^{-1}$. We apply periodic boundary conditions in all three spatial directions, ensuring that $f\left(\mathbf{n}+\mathbf{e}_iN\right)=f\left(\mathbf{n}\right)$, where $\mathbf{e}_i$ denotes one of the unit vectors in the three spatial dimensions.

Note that the finite volume of the cubic lattice implies an IR cut-off of momenta
\begin{equation}
	\delta k=\frac{2\pi}{N\delta x}\ ,
\end{equation}
and therefore the momenta must be discretized. The finite spatial volume and the discretization require discrete Fourier transformation,
\bea
f\left(\Tilde{\bf n}\right)&=&\sum_{\mathbf{n}}\left(\delta x\right)^3f\left(\mathbf{n}\right)\e^{-\text{i}\frac{2\pi}{N}\mathbf{n}
	\cdot\Tilde{\mathbf{n}}}, \nnn f\left(\mathbf{n}\right)&=&\sum_{\Tilde{\mathbf{n}}}\left(\frac{\delta k}{2\pi}\right)^3f\left(\Tilde{\mathbf{n}}\right)\e^{\text{i}\frac{2\pi}{N}\Tilde{\mathbf{n}}\cdot\mathbf{n}}\ ,
\eea
where the momenta are also periodic, so practically we choose
\begin{equation}
	\Tilde{\mathbf{n}}=\left(\Tilde{n}_1,\Tilde{n}_2,\Tilde{n}_3\right)\ ,\ \text{with}\ \Tilde{n}_i=-\frac{N-1}{2},\dots,\frac{N-1}{2}\ .
\end{equation}
and the corresponding comoving momenta are
\begin{equation}
	\mathbf{k}=\left(k_1,k_2,k_3\right),\ \text{with}\ k_i=-\frac{N-1}{N}\frac{\pi}{\delta x},\dots,\frac{N-1}{N}\frac{\pi}{\delta x}\ .
\end{equation}

\begin{figure}[htpb]
	\centering
	\includegraphics[width=\linewidth]{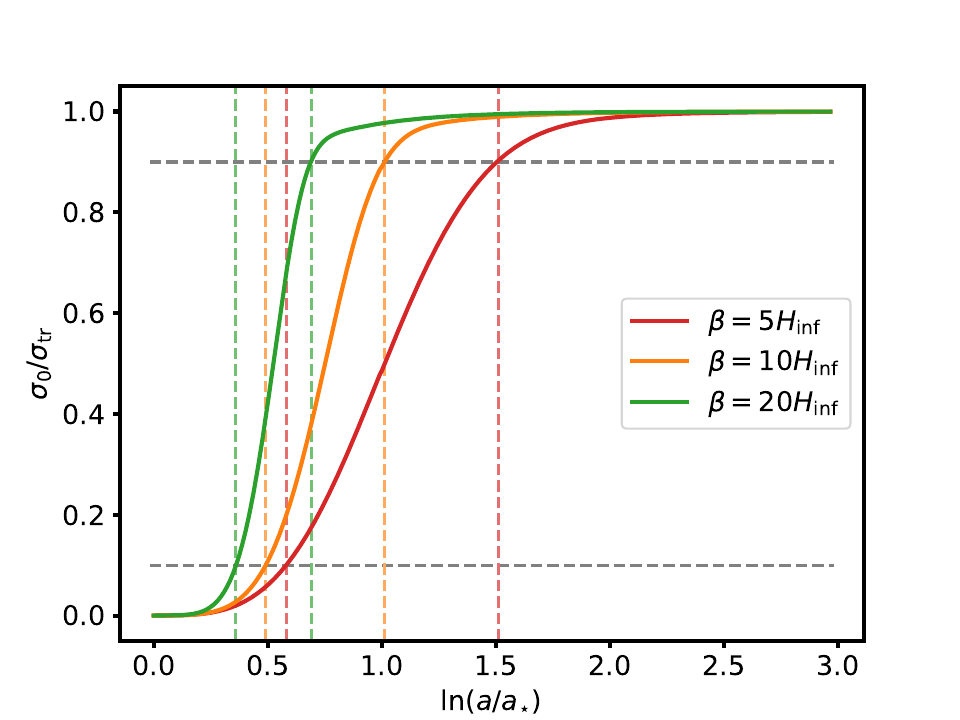}
	\caption{The mean value of $\sigma$. The dashed horizontal lines label the situation $10\%$ (down) or $90\%$ (up) of the space is occupied by the true vacuum while the dashed vertical lines label the moment at which the system reach such situations.   }  \label{fig:sigma}
\end{figure}

For time evolution, we set the scale factor $a\left(\tau_\star\right)=1$ at the starting time of the simulation, so the conformal time $\tau_\star=-H_{\rm inf}^{-1}$. To ensure accurate temporal resolution, we choose different temporal steps for each e-fold. In practice, the phase transition typically completes within a maximum of 3 e-folds, as shown in Fig.~\ref{fig:sigma}. We divide the first one into 1000 steps, the second one into 500 steps, and the final one into 250 steps.

To facilitate the numerical simulation, we redefine the fields and parameters in the theory as
\begin{equation}
	\sigma\rightarrow\frac{\sigma}{\sqrt{\lambda}},\ \phi\rightarrow \frac{\phi}{\sqrt{c_m}},\ c_3\rightarrow\sqrt{\lambda}c_3\ .
\end{equation}
Then Eq.~(\ref{V}) becomes
\begin{equation}
	V_1\left(\phi,\sigma\right)=\frac{1}{\lambda}\left[-\frac{1}{2}\left(1-\phi^2\right)\sigma^2+\frac{1}{3}c_3\sigma^3+\frac{1}{4}\sigma^4\right]\ ,
\end{equation}
and the field equation Eq.~(\ref{eq:sigma}) can be rewritten as
\begin{equation}
	\sigma''+2\mathcal{H}\sigma'-\nabla^2\sigma+a^2\left[\left(\phi^2-1\right)\sigma+c_3\sigma^2+\sigma^3\right]=0\ .
\end{equation}

The starting time of simulation, $t_\star$, is defined as the moment when approximately one bubble is nucleated per Hubble volume, represented by the condition
\begin{equation}
	\frac{\Gamma(t_\star)}{V_{\rm phys}}\simeq H_{\rm inf}^4\ .
\end{equation}
During the phase transition, the bubble nucleation rate can be parameterized as
\begin{equation}
	\Gamma(t)\simeq \Gamma(t_\star) \e^{\beta(t-t_\star)}\ ,
\end{equation}
where
\begin{equation}
	\beta \equiv-\left.\frac{d S_b}{d t}\right|_{t=t_\star}\ .
\end{equation}
Here $t$ denotes physical time. 

To ensure the phase transition can take place and complete within a reasonable time frame, the action $S_{\text{b}}$ should not be too large, which we keep under 20. Under this requirement and the slow-roll condition, we set the quartic interaction coefficient $\lambda=2$, the cubic coefficient $c_3=1$, the initial value of the inflaton $\phi\left(t_\star\right)=1.03$, and then adjust the initial velocity of the inflaton $\dot\phi\left(t_\star\right)$ to tune the value of $\beta$.

As demonstrated in Eq.~(B18) of [84] and Eq.~(A8) of [85], for an first-order phase transition during inflation, we have
\begin{equation}\label{eq:betaoverH}
	\frac{\beta}{H_{\rm inf}} \approx I_1 S_{\text{b}} \times \frac{1}{N_e \left|  1 - \frac{1}{\phi^2} \right|} \ ,
\end{equation}
where $I_1$ is an $\mathcal{O}\left(1\right)$ parameter, $N_e$ is the number of e-folds before the end of inflation. It is evident that in comparison to the typical phase transition during the thermal big bang expansion, the value $\beta/H_{\rm inf}$ is suppressed by a factor of $1/N_e$. Generally, the combination $\left|  1 - \frac{1}{\phi^2} \right|$ is also a $\mathcal{O}\left(1\right)$ parameter, but its detailed value is model dependent. And $S_{\text{b}}$ is about $\mathcal{O}\left(100\right)$. In the scenario of high scale inflation, the total number of e-folds when the CMB modes exit the horizon is approximately 60. Therefore, $N_e$ in Eq.~(\ref{eq:betaoverH}) is about 45. Consequently, in this model, it is natural for $\beta/H_{\rm inf}$ to be less than $\mathcal{O}(10)$.

The initial condition of $\sigma$ is chosen as
\begin{equation}
	\sigma=\sigma_{\rm fl},\quad \sigma'=0\ .
\end{equation}
we simulate the phase transition process using the classical equation of motion for $\sigma$. The evolution of this field is computed using the same numerical integrator employed in~\cite{Forest:1989ez,Yoshida:1990zz}, (the details can also be found in the appendix of~\cite{An:2023idh}). 
In the case of this work, the Hamiltonian governing the system's evolution is
\begin{equation}
	\mathcal{H}=\int d^3\mathbf{x} \frac{1}{2}\left[\frac{1}{a^2}\pi^2+a^2(\nabla \sigma)^2+a^4 V_1\left(\phi,\sigma\right)\right]\ ,
\end{equation}
where $\pi\equiv a^2 \sigma'$. 

During the simulation, the bubbles are generated within the region still occupied by the false vacuum at the beginning of each temporal step. The number density of bubbles generated at each temporal step is calculated using the nucleation rate per unit lattice volume, and the positions of bubble centers are randomly selected.

At each temporal step, the probability for a bubble to be produced at a given lattice site follows a binomial distribution with the probability, 
\begin{equation}
	p = \frac{\Gamma(\tau)}{V_{\rm phys}}\delta x^3\delta \tau a^4(\tau)\ .
\end{equation}
Practically, during each temporal step, random numbers are generated according to the binomial distribution with probability $p$ at the lattice sites that are still occupied by the false vacuum. These random numbers determine whether true vacuum bubbles can be generated at each respective site. The profile of the bubbles is determined by the bounce solution $\sigma_b$. 

After nucleation, the bubbles expand rapidly, collide with each other and eventually occupy the entire space. Fig.~\ref{fig:sigma} illustrates the evolution of  ratio $\sigma_0/\sigma_{\rm tr}$,  where  $\sigma_0$ is the mean value of the $\sigma$ field. 
This ratio reflects the occupation fraction of the true vacuum, as the false vacuum of our system is set to $\sigma_{\rm fl}=0$. Fig.~\ref{fig:sigma} demonstrates that the bubble collision process is completed within a few $\beta^{-1}$.

With the $\sigma$ field configuration at each temporal step, we calculate $\delta\phi$, $\Psi$ and $\Phi$ by solving the differential equations (6) and (9) together with the conditions Eqs.~(7), (10) and (11). 

Note that $\delta \phi$ contributes to the source term of $\Psi$ in Eq.~(9), so we cannot directly calculate the late-time limit of $\delta\phi$ using the Green's function method described in the main text for the qualitative analysis. In the numerical simulation, the differential equations are solved iteratively. Specifically, at each temporal step, we use the values of $\sigma$, $\delta\phi$, $\Psi$ and $\Phi$ from the previous step to calculate the source functions Eqs.~(7), (9) and (11). We then use these source functions to calculate the evolution of $\delta\phi$, $\Psi$, and $\Phi$ in the current step. This iterative calculation is performed to obtain the evolution of $\delta\phi$, $\Psi$, and $\Phi$. 

\begin{figure}[htpb]
	\centering
	\includegraphics[width=1\linewidth]{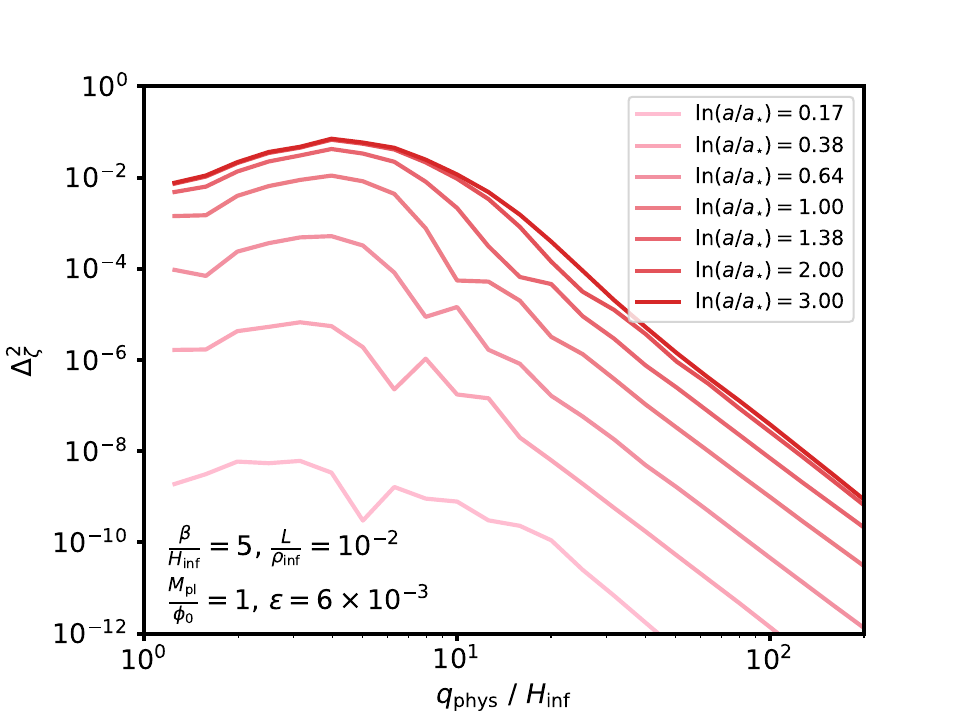}
	\caption{Accumulated contributions to the induced curvature perturbation, with the $\tau'$ integral stops at the values shown in the legend for each curve.}\label{fig:accumulated}
\end{figure}

After the phase transition, we calculate the curvature perturbation $\zeta$ using Eq.~(12), as well as the power spectrum $\Delta_\zeta^2$. Fig.~\ref{fig:accumulated} illustrates the accumulated contributions to $\Delta_\zeta^2$, where each curve represents the integral over $\tau'$ from $\tau_\star$ to the indicated value in the figure. The parameters chosen for the simulation are $\beta/\hinf = 5$, $L/\rho_{\rm inf} = 10^{-2}$, $\mpl/\phi_0 = 1, \epsilon = 6\times10^{-3}$. We can see that the induced curvature perturbations are mostly generated between one and two e-folds after the phase transition. Therefore, the physical duration of the curvature perturbation production is approximately $\hinf^{-1}$. 

\begin{figure}[htpb]
	\centering
	\includegraphics[width=\linewidth]{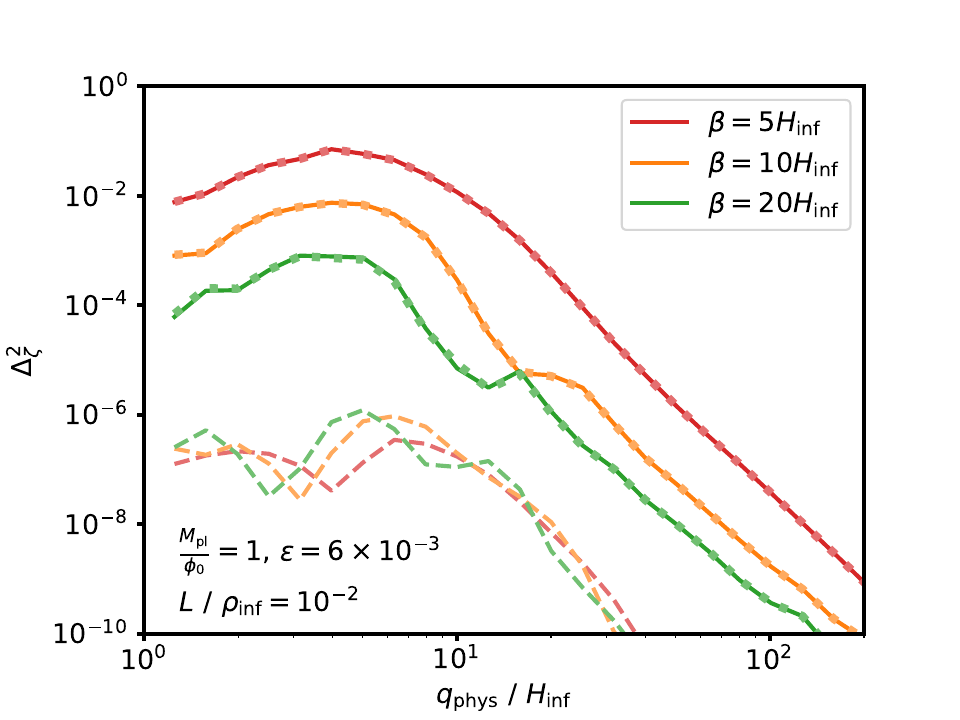}
	\caption{The power spectra of curvature perturbation for different $\beta$. The dotted line shows the contribution from the direct coupling and the dashed line shows the pure gravitational contribution, while the full results are shown by the solid line. We can observe the solid line and the dotted line are almost overlapping. }  \label{fig:Delta12}
\end{figure}

Fig.~\ref{fig:Delta12}  displays the induced curvature spectrum $\Delta_\zeta^2$ for various choices of $\beta/\hinf$, along with the dashed curves representing the pure gravitational contributions. It is evident that the gravitational contributions are negligible compared to the direct contributions.

\section{Details of the shape function $\mathcal{F}_2$ }

\begin{figure}[htpb]
	\centering
	\includegraphics[width=0.9\linewidth]{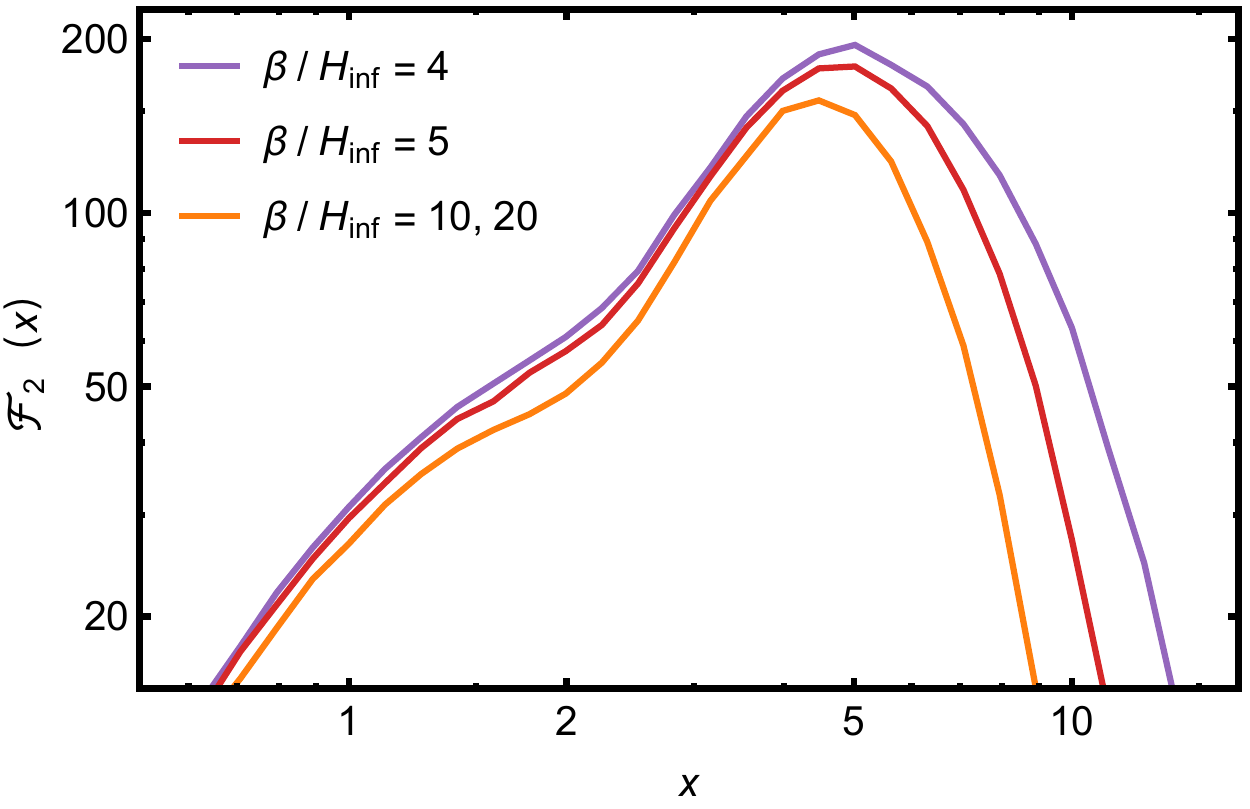}
	\caption{The shape function $\mathcal{F}_2(x)$ of different $\beta$.}  \label{fig:F2}
\end{figure}

For the curvature perturbation to reenter the horizon well before the matter-radiation equality, the GWs observed today are mainly generated during RD. Thus the energy fraction of GWs can be written as~\cite{Kohri:2018awv}
\begin{equation}
	\begin{split}
		\Omega_{\text{GW}}(k)&=\Omega_{\text{rad}}\frac{k^3}{6} \int_0^{\infty}d k_1\int_{-1}^{1}d\mu \frac{k_1^3}{k_2^3}\left(1-\mu^2\right)^2\\
		&\cdot  \overline{ I^2}(k,k_1,k_2) \mathcal{P}_{\zeta}\left(k_1\right)\mathcal{P}_{\zeta}\left(k_2\right)\ ,
	\end{split}
\end{equation}
where $\overline{ I^2}(k,k_1,k_2)$ is the integration kernel
\bea 	 
\overline{ I^2}(k,k_1,k_2)&=& \frac{1}{2}\left(\frac{3(k_1^2+k_2^2-3k^2)}{4k_1^3k_2^3}\right)^2 \nnn 
&&\!\!\!\!\!\!\!\!\!\!\!\!\!\!\!\!\!\!\!\!\!\!\!\!\!\!\!\!\!\!\!\!\!\!\!\!\!\!\!\!\!\!\!\!\times\Bigg\{\pi^2\left(k_1^2+k_2^2-3k^2\right)^2\theta\left(k_1+k_2-\sqrt{3}k\right) \nnn
&&\!\!\!\!\!\!\!\!\!\!\!\!\!\!\!\!\!\!\!\!\!\!\!\!\!\!\!\!\!\!\!\!\!\!\!\!\!\!\!\!\!\!\!\!+\left[-4k_1k_2+\left(k_1^2+k_2^2-3k^2\right)\log\left|\frac{3k^2-\left(k_1+k_2\right)^2}{3k^2-\left(k_1-k_2\right)^2}\right|\right]^2\Bigg\}\ . \nnn
\eea
with $\mathbf{k}=\mathbf{k}_1+\mathbf{k}_2$ and $\mu=\frac{\mathbf{k} \cdot \mathbf{k}_1}{k k_1}$. It's convenient to introduce a new variable $v=k_1/k$, then we have
\begin{equation}
	\begin{split}
		\Omega_{\text{GW}}(k)=&\Omega_{\text{rad}}\int_0^{\infty}d v\int_{-1}^{1}d\mu \left(1-\mu^2\right)^2\mathcal{K}(v,\mu)\\
		&\cdot\mathcal{P}_{\zeta}\left(v k\right)\mathcal{P}_{\zeta}\left(\sqrt{v^2+1-2v\mu}k\right)\ ,
	\end{split}
\end{equation}
where
\begin{equation}
	\mathcal{K}(v,\mu)\equiv \frac{v^3}{6}\frac{\overline{ I^2}\left(1,v,\sqrt{v^2+1-2v\mu})\right)}{(v^2+1-2v\mu)^{3/2}}\ .
\end{equation}
Notice the asymptotic behaviors of $\mathcal{K}(v,\mu)$ is well approximated by
\begin{equation}
	\mathcal{K}(v,\mu)\simeq \left\{
	\begin{matrix}
		v^3/3 & v\ll1\\
		3v^{-4}\log^2v & v\gg1\\
	\end{matrix}\right.\ ,
\end{equation}
which is independent of $\mu$.
The shape function $\mathcal{F}_2$ defined in the main text can be calculated by
\begin{equation}
	\begin{split}
		\mathcal{F}_2(x)=&\int_0^{\infty}d v\int_{-1}^{1}d\mu \left(1-\mu^2\right)^2\mathcal{K}(v,\mu)\\
		&\cdot \mathcal{F}\left(v x\right)\mathcal{F}\left(\sqrt{v^2+1-2v\mu}x\right)\ ,
	\end{split}
\end{equation}
where
\begin{equation}
	\mathcal{F}(x)=\frac{x^3}{1+(\alpha_1 x)^4+(\alpha_2 x)^9}
\end{equation}
is the form factor for $\Delta_\zeta^2$.
We assume $\mathcal{F}(x)$ has a maximum value $\mathcal{F}_{\rm max}$ with coordinate $x_{\rm max}$, and it can be approximated as
\begin{equation}
	\mathcal{F}(x)\simeq \left\{
	\begin{matrix}
		\mathcal{F}_{\rm max}(x/x_{\rm max})^3& x \ll x_{\rm max}\\
		\mathcal{F}_{\rm max}(x/x_{\rm max})^{-6}& x \gg x_{\rm max}\\
	\end{matrix}\right.\ .
\end{equation}
Considering the asymptotic behaviors of $\mathcal{F}_2$, the integral over $v$ is predominantly contributed in the vicinity of  $vx\simeq x_{\rm max}$, which corresponds to the point where $\mathcal{F}(x)$ reaches its maximum value, since the descent of $\mathcal{F}(x)$ is steeper compared to the behavior of  $\mathcal{K}(v,\mu)$. For small $x$, $v\simeq x_{\rm max}/x\gg1$, $\sqrt{v^2+1-2v\mu}\simeq v$, then the integral becomes
\begin{equation}
	\begin{split}
		\mathcal{F}_2(x)\simeq&\mathcal{F}_{\rm max}^2\Bigg\{\int_0^{x_{\rm max}/x}dv 3v^{-4}\log^2 v \cdot (vx/x_{\rm max})^6\\
		&+\int_{x_{\rm max}/x}^\infty dv 3v^{-4}\log^2 v \cdot (vx/x_{\rm max})^{-12}\Bigg\}\\
		\simeq&\mathcal{F}_{\rm max}^2(x/x_{\rm max})^3\log^2(x/x_{\rm max})\ ,
	\end{split}
\end{equation}
where the $\mu$ integral is factorized and the result is a $\mathcal{O}(1)$ number. And for large $x$, $v\simeq x_{\rm max}/x\ll1$, $\sqrt{v^2+1-2v\mu}\simeq 1$. Similarly, we finish the integration and get
\begin{equation}
	\begin{split}
		\mathcal{F}_2(x)\simeq&\mathcal{F}_{\rm max}^2\Bigg\{\int_0^{x_{\rm max}/x}dv v^3/3 \cdot (vx/x_{\rm max})^3 (x/x_{\rm max})^{-6}\\
		&+\int_{x_{\rm max}/x}^\infty dv v^3/3 \cdot (vx/x_{\rm max})^{-6}(x/x_{\rm max})^{-6}\Bigg\}\\
		\simeq&\mathcal{F}_{\rm max}^2(x/x_{\rm max})^{-10}\ .
	\end{split}
\end{equation}
In the parameter space of interest, we have $\mathcal{F}_{\rm max}\simeq \mathcal{O}(10)$, resulting in a peak value of $\mathcal{F}_2(x)$ at around $\mathcal{O}(100)$. The shape of $\mathcal{F}_2(x)$ for $\beta/H_{\rm inf}=4,5,10,20$ is depicted in Fig.~\ref{fig:F2}. We observe a shoulder on the left side of the peak, which arises from the contribution of logarithmic function. This shoulder significantly slow down the slope in that region.

\section{Details of the Bayesian analysis}

In the Bayesian data analysis, we follow the same approach as described in \cite{NANOGrav:2023hvm}. Here, we review the main idea. The signal of new physics is encoded in the timing residuals of pulsars, denoted as $ \vec{\delta t}$, which can be divided into three components: white noise, red noise, and small errors in the fit to the timing-ephemeris parameters \cite{NANOGrav:2020tig}. The red noise consists of both pulsar-intrinsic red noise and GW signals. We denote all possible parameters as $\vec{\theta}$. 

The PTA likelihood function describes the probability of observing the data $ \vec{\delta t}$ given the parameter values $\vec{\theta}_0$, denoted as $p( \vec{ \delta t}|\vec{\theta}_0)$. To obtain the marginalized likelihood for the parameters of interest, we integrate over the other parameters. The posterior probability distribution for the model parameters $\vec{\theta}_0$ is given by Bayes' theorem
\begin{equation}
	p( \vec{\theta}_0 |\vec{\delta t} )=\frac{p( \vec{ \delta t}|\vec{\theta}_0)p(\vec{\theta}_0)}{p( \vec{ \delta t})}.
\end{equation}
For model comparison, the Bayes factor describes whether a model $\mathcal{M}$ is favored or disfavored by the data $ \vec{\delta t}$, which is defined as 
\begin{equation}
	\mathcal{B}=\frac{p( \vec{ \delta t}|\mathcal{M})}{p( \vec{ \delta t}|\mathcal{M}_0)},
\end{equation}
where in our case $\mathcal{M}_0$ is the reference model in which the GWs are only sourced by SMBHB. To place a bound to a parameter $\theta$, we utillize the K-ratio
\begin{equation}
	K(\theta)=\frac{p( \vec{ \delta t}|\mathcal{M},\theta)}{p( \vec{ \delta t}|\mathcal{M},\bar{\theta})},
\end{equation}
where $\bar{\theta}$ refers to the limit in which $p( \vec{ \delta t}|\mathcal{M},\theta)$ no longer depends on the value of $\theta$.

\begin{table}[htbp]
	\vspace{4ex}
	\centering
	\begin{tabular}{ccc}
		\hline \noalign{\smallskip}
		Parameter & Description & Prior\\
		\hline \noalign{\smallskip}
		$A_{\rm ref}$ & reference amplitude & log-uniform[-10,-2]\\
		\noalign{\smallskip}
		$f_{\rm ref}$ & reference frequency & log-uniform[-5,4]\\
		\hline \noalign{\smallskip}
	\end{tabular}
	\caption{Priors distributions for the parameters.}\label{tab:1}
\end{table}

In our analysis, we utilize the software \textsf{PTArcade}~\cite{andrea_mitridate_2023,Mitridate:2023oar}, which is based on several libraries including \textsf{ceffyl}~\cite{lamb2023need}, \textsf{ENTERPRISE}~\cite{enterprise}, \textsf{ENTERPRISE-EXTENSIONS}~\cite{enterprise-ext}, \textsf{PTMCMCSampler}~\cite{justin_ellis}, \textsf{corner}~\cite{Foreman-Mackey2016} and \textsf{GetDist}~\cite{Lewis:2019xzd}. We employ \textsf{PTArcade} to calculate the PTA likelihood and the Bayes factor.

In this work, the phase-transition induced curvature perturbation depends on $A_{\rm ref}$, $f_{\rm ref}$, and $\hinf/\beta$. For the the Bayes factor analysis, we consider two cases, $\beta=5 H_{\rm inf}$ and $\beta=10,\ 20 H_{\rm inf}$.  Then, the power spectra can be written as
\begin{equation}
	\Delta^2_\zeta(f)=A_{\rm ref}\frac{(f/f_{\rm ref})^3}{1 + (0.31 f/f_{\rm ref})^4 + (0.17 f/f_{\rm ref})^{9}},
\end{equation}
and
\begin{equation}
	\Delta^2_\zeta(f)=A_{\rm ref}\frac{(f/f_{\rm ref})^3}{1 + (0.31 f/f_{\rm ref})^4 + (0.2 f/f_{\rm ref})^{9}},
\end{equation}
respectively.

The priors distributions for the parameters are provided in Tab.~\ref{tab:1}. For the red noise, we employ 30 frequency bins for pulsar-intrinsic red noise and 14 frequency bins for GW backgrounds, as suggested in \cite{Mitridate:2023oar}. In the Bayesian analysis, we use the bootstrapping method outlined in \cite{NANOGrav:2023hvm}. We obtain a Bayes factor $\mathcal{B}=153.0\pm0.0$ for $\beta=5 H_{\rm inf}$ and $\mathcal{B}=272.0\pm0.0$ for $\beta=10,\ 20 H_{\rm inf}$. We use the criterion $K>1/10$ to establish the K-ratio bound for our parameters which is shown by the gray region in Fig.~4. The main results of Bayesian analysis are listed in Tab.~\ref{tab:2} and Tab.~\ref{tab:3} while the K-ratio bound is not reached for both two individual parameters.

\begin{table}[htbp]
	\centering
	\begin{tabular}{cccc}
		\hline \noalign{\smallskip}
		Parameter & \makecell{Bayes \\ Estimator} & \makecell{Maximum \\ Posterior} & \makecell{68\% Credible \\ Interval} \\
		\hline \noalign{\smallskip}
		$\log_{10}A_{\rm ref}$ & -1.67$\pm$1.29 & -2.35 & [-2.86,-1.66]\\
		\noalign{\smallskip}
		$\log_{10}f_{\rm ref}$ &-7.25$\pm$1.04 & -7.78& [-8.28,-7.09]\\
		\hline \noalign{\smallskip}
	\end{tabular}
	\caption{Bayesian estimators, maximum posterior values, and $68\%$ credible intervals for the parameters $A_{\rm ref}$ and $f_{\rm ref}$ for $\beta=5 H_{\rm inf}$.}\label{tab:2}
\end{table}

\begin{table}[htbp]
	\centering
	\begin{tabular}{cccc}
		\hline \noalign{\smallskip}
		Parameter & \makecell{Bayes \\ Estimator} & \makecell{Maximum \\ Posterior} & \makecell{68\% Credible \\ Interval} \\
		\hline \noalign{\smallskip}
		$\log_{10}A_{\rm ref}$ & -1.68$\pm$1.20 & -2.33 & [-2.82,-1.64]\\
		\noalign{\smallskip}
		$\log_{10}f_{\rm ref}$ &-7.26$\pm$0.98 & -7.79& [-8.26,-7.09]\\
		\hline \noalign{\smallskip}
	\end{tabular}
	\caption{Same as in Tab.~\ref{tab:2} but for $\beta=10,\ 20 H_{\rm inf}$.}\label{tab:3}
\end{table}

\end{document}